\documentclass[12pt]{article}
\usepackage[colorlinks=true,linkcolor=blue,urlcolor=blue,filecolor=black,citecolor=red,pdfstartview=FitV,pdftitle={},pdfsubject={},pdfkeywords={},pdfpagemode=None,bookmarksopen=true]{hyperref}
\usepackage{graphicx}
\usepackage{epstopdf}%
\usepackage{amsmath}
\usepackage{amsfonts}
\usepackage{amssymb}
\usepackage{color}%
\usepackage{dcolumn}
\usepackage{slashed}
\usepackage{amssymb,ulem}
\usepackage{float}
\usepackage{amsthm,amsmath,amssymb}
\usepackage{mathrsfs}
\usepackage{subfigure}
\usepackage{wrapfig}
\usepackage{indentfirst}
\providecommand{\U}[1]{\protect\rule{.1in}{.1in}}

\newcommand{\f}{\begin{equation}}
\newcommand{\ff}{\end{equation}}
\newcommand{\fa}{\begin{eqnarray}}
\newcommand{\ffa}{\end{eqnarray}}

\begin{document}
\title{Gravitational decoupling for hairy black holes in asymptotically AdS spacetimes }
\author{Chao-Ming Zhang$^{1}$ \thanks{\href{mailto:zcm843395448@163.com}
        {zcm843395448@163.com}},
       Ming Zhang$^{2}$ \thanks{\href{mailto:mingzhang0807@126.com}
        {mingzhang0807@126.com}},
         and De-Cheng Zou$^{1}$ \thanks{Corresponding author. \href{mailto:dczou@yzu.edu.cn}
        {dczou@yzu.edu.cn}}\\
$^{1}$Center for Gravitation and Cosmology, College of Physical Science \\  and Technology,
Yangzhou University, Yangzhou 225009, China\\
$^{2}$Faculty of Science, Xi'an Aeronautical University,\\ Xi'an 710077, China}
\date{}
\maketitle

\begin{abstract}
	In this paper, the gravitational decoupling approach via extended geometric deformation (EGD) is utilized
	to generate analytical black hole solutions owing to its simplicity and effectiveness.
	Considering the external fields surrounding Schwarzschild AdS black hole, we derive
	hairy black hole solutions in asymptotically AdS spacetime, satisfying
	the strong energy and dominate energy conditions, respectively.
	Moreover, we find that if the black hole spacetime is a fluid system,
	the fluid under each of these conditions is anisotropic.

\end{abstract}

\section{Introduction}\label{int}
\par
In general relativity (GR), there is a theorem about the properties of black holes called ``no-hair theorem" \cite{Carter:1971zc}-\cite{Chrusciel:2012jk},
which states that no more than three characteristic parameters affect black holes: mass, angular momentum and charge.
However, in recent years, with the deepening of the exploration of the universe and black holes, it has been
discovered that there may be other physical quantities that affect the properties of black holes.
For example, black holes could contain quantum hairs~\cite{Hawking:2016msc} when there
exist other charges associated with inner gauge symmetries (and fields), and black holes with scalar hair were also
discussed in scalar-tensor theories~\cite{Sotiriou:2011dz}-\cite{Antoniou:2017acq}, where the scalar fields play a pivotal
role in particle physics and early universe cosmology~\cite{Svrcek:2006yi,Peebles:1998qn}.
With the recent release of black hole images and the discovery of gravitational waves, the scope to test
the no-hair theorem has been further enhanced~\cite{Berti:2013gfa}.
Nevertheless, it's usually very difficult to find analytical hairy solutions describing general matter configurations.

It is worth noting that a new and elegant method--gravitational decoupling that allows us
to obtain new exact solutions starting from a known one has received considerable attention recently.
This approach by means of minimal geometric deformation (MGD) was developed to
deform Schwarzschild spacetime in the context of braneworld~\cite{Ovalle:2009xk,Ovalle:2016pwp}.
Subsequently, many applications of the method and its extension (hereafter referred to as EGD) have been studied for the stellar systems
in the braneworld framework~\cite{Ovalle:2007bn}-\cite{Ovalle:2015nfa}.
MGD is a well succeeded theory that allows the study of nonlinear gravity in braneworlds
to obtain the effective action at low energies by using the AdS/CFT correspondence~\cite{Kanno:2002iaa,Soda:2010si}.

Beyond the braneworld, this gravitational decoupling approach has also been used to generate solutions of Einstein
field equations in GR and other modified gravitational theories, since it has following advantages: (a) it
can decouple the complex energy-momentum tensor into various relatively simpler components; (b) one can use
it to extend some known seed solutions to more complex solutions; (c) one can use it to find solutions
of gravitational theories other than GR. The procedure for gravitational decoupling approach can be expressed
as follows: ``Suppose there are two non-interacting gravitational
sources in a spacetime: a seed source $T_{\mu\nu}$ and an extra source $\theta_{\mu\nu}$.
The standard Einstein’s equation are first solved for $T_{\mu\nu}$, and then another
set of quasi-Einstein equations are solved for $\theta_{\mu\nu}$. Subsequently, the two solutions can
be combined in order to derive the complete solution for the total system''.
Mathematically, the total energy-momentum tensor of the gravitational system is given by
\begin{eqnarray}
\tilde{T}_{\mu\nu}=T_{\mu\nu}+\theta_{\mu\nu}.
\end{eqnarray}
Here, $\theta _{\mu\nu}$ may be a scalar field, a vector field, etc.
By applying the algorithm to previously known seed solutions, this method has adopted to derive several
new solutions of gravitational systems in GR, as for example stellar
distributions \cite{Ovalle:2019lbs,Morales:2018nmq,Torres-Sanchez:2019wjv} and
black holes \cite{Ovalle:2018umz}. Through the gravitational decoupling approach,
Ovalle derived the anisotropic solutions of Einstein field equations for self-gravitating systems from
perfect fluid solutions by using MGD technique~\cite{Ovalle:2017fgl}. More recently, gravitational decoupling via MGD was used
to investigate higher dimensional compact structures~\cite{Estrada:2018vrl} and spread out to the context of
Horndeski~\cite{daRocha:2021sqd}, Lovelock (Gauss-Bonnet)~\cite{Estrada:2019aeh}-\cite{Maurya:2022uqu},
Rastall~\cite{Maurya:2019xcx} and $f(R,T)$~\cite{Maurya:2019iup}
gravity theories and the cosmological scenario~\cite{LinaresCedeno:2019aul}.

However, the MGD technique has some limitations because it only deforms the radial metric potential
by leaving temporal coordinate as an invariant quantity. It results in that the considered sources do not
exchange energy in the scenario. Fortunately, the extension version of MGD (EGD henceforth) was presented
in~\cite{Ovalle:2018gic}, where J. Ovalle considered the decoupling of two spherically symmetric and static
gravitational sources in GR and found the exchange of energy between decoupled sources.
Moreover, the corresponding metrics possess a well-defined event horizon and also reproduced
the Reissner-Nordstrom solution.
M.~Sharif \textit{et al.}\cite{Sharif:2020iui,Sharif:2020vvk} obtained astrophysical
and cosmological solutions in the context of SBD theory by using MGD and EGD techniques, respectively.
E.~Contreras~\textit{et al.}\cite{Contreras:2019mhf} successfully decoupled the field equations in $(1+2)$-dimensional gravity
to obtain exterior charged BTZ model from the corresponding vacuum solution through EGD technique.

Now, Let $g_{\mu\nu}$ be any metric. After calculating its Einstein tensor $G_{\mu\nu}$, $g_{\mu\nu}$ can be
regarded as the metric corresponding to the dynamic tensor $T_{\mu\nu} = \frac{G_{\mu\nu}}{8 \pi}$.
The problem is that such arbitrary $T_{\mu\nu}$ is not necessarily a very dirty amount of motion of some material field.
Therefore, such a `` solution " is not necessarily physically meaningful.
It is generally believed that the dynamic tensor of any material field must satisfy a number of ``reasonable" conditions,
collectively known as the energy conditions~\cite{liang}. The energy conditions imposed on the energy-momentum tensor
can be generally considered as sensible guidelines to avoid classically unphysical configurations
and exotic matter sources apart from the Einstein field equations in the context of a
wide class of spacetime theories~\cite{Curiel:2014zba}.
Considering the deformation of seed Schwarzschild vacuum spacetime, Ovalle \textit{et al.}\cite{Ovalle:2020kpd} recently
formulated new hairy black holes through the EGD technique for the first time, where the additional source is described by a conserved
energy-momentum tensor $\theta_{\mu\nu}$  which satisfies either the strong (SEC) or dominant energy condition (DEC)
in the region outside the event horizon. Subsequently, hairy black holes were discussed from the perspective
of conformal anomalies~\cite{Meert:2021khi}. From a spherically symmetric seed solution, the solutions of rotating
hairy black hole were also constructed using the EGD technique~\cite{Contreras:2021yxe,Mahapatra:2022xea},
including the studies of strong field gravitational lensing effects~\cite{Islam:2021dyk}
and thermodynamic properties~\cite{Mahapatra:2022xea} of these black holes.
It is well known that the AdS/CFT duality generally states that weakly-coupled gravity in $(d+1)$-dimensional
anti-de Sitter (AdS) space is the theory dual to a strongly-coupled conformal field theory (CFT).
It is interesting to explore the applications for gravitational decoupling approach in asymptotically AdS spacetimes.
Notice that some new solutions of Einstein field equations in asymptotically (A-)dS spacetimes were also presented
in Refs.~\cite{Contreras:2019fbk}-\cite{Contreras:2018nfg} using MGD technique recently.
Moreover, the AdS/CFT correspondence setup has been employed to compute the Bose-Einstein condensates (BCS) \cite{Fernandes-Silva:2019fez}
and holographic entanglement entropy (HEE) corrections \cite{DaRocha:2019fjr,daRocha:2020gee}
for spherically symmetric spacetimes in the context of MGD and EGD.
Retaining the energy conditions, here we will adopt EGD technique to derive the solutions of hairy black hole
in asymptotically AdS spacetimes. We expect that the obtained hairy AdS black holes can be investigated
to explore aspects of gravity in the strong nonlinear regime and comparing any deviation from the GR setup
in the gravitational wave astrophysical scenario.

The paper is organized as follows. In Sec.~\ref{es}, we review the gravitational decoupling approach
via EGD in asymptotically AdS spacetime.
In Sec.~\ref{bh}, we separately adopt the SEC and DEC to derive two new
families of hairy black hole solutions. We summarized the results of our study in Sec.~\ref{con}.

\section{Gravitational decoupling approach} \label{es}

We consider the Einstein field equations
\begin{eqnarray}
G_{\mu\nu}=R_{\mu\nu}-\frac{1}{2}R g_{\mu\nu}+\Lambda g_{\mu\nu}=\kappa^2 \tilde{T}_{\mu\nu},\label{eqeins}
\end{eqnarray}
and assume that the total energy-momentum tensor is
given by
\begin{eqnarray}
\tilde{T}_{\mu\nu}=T_{\mu\nu}+\theta_{\mu\nu},\label{theta1}
\end{eqnarray}
where $\Lambda$ is the negative cosmological constant and $T_{\mu\nu}$ is the 4-dimensional energy-momentum
tensor of a perfect fluid $T_{\nu}^{~\mu}=diag(-\rho,p,p,p)$.
Moreover, $\theta_{\mu\nu}$ describes an additional source
whose coupling to gravity is proportional to the constant $\alpha$.
Since the Einstein tensor $G_{\mu\nu}$ is divergence free, the total energy-momentum tensor
$\tilde{T}_{\mu\nu}$ satisfies the conservation equation
\begin{eqnarray}
{\nabla _\nu }{\tilde{T}^{\mu \nu }} = 0.
\end{eqnarray}

From the expression of energy-momentum tensor in Eq.~\eqref{theta1},
we define an effective density with
\begin{eqnarray}
\tilde{\rho} =-\tilde{T}_0^{~0}=\rho-\theta _0^{~0},
\end{eqnarray}
the effective radial pressure
\begin{eqnarray}
\tilde{p}_r=\tilde{T}_1^{~1}=p_r+\theta _1^{~1},
\end{eqnarray}
and the effective tangential pressure
\begin{eqnarray}
\tilde{p}_{\bot}=\tilde{T}_2^{~2}=p_{\bot}+\theta _2^{~2}.
\end{eqnarray}

In this study, we investigate the static and spherically symmetric metric
\begin{eqnarray}
d{s^2}=-{e^{\nu (r)}}d{t^2}+{e^{\lambda (r)}}d{r^2}+{r^2}(d{\theta ^2}+{\sin ^2}\theta d{\phi ^2}),\label{eqmetrictotal}
\end{eqnarray}
where $\nu =\nu(r)$ and $\lambda=\lambda(r)$ are functions of the areal radius $r$ only.
Substituting the above metric into Einstein field equations (Eq.(\ref{eqeins})), we get
\begin{eqnarray}
\kappa ^2 \left(-\rho+\theta _0^{~0}\right)&=&-\frac{1}{r^2}+e^{-\lambda } \left(\frac{1}{r^2}
-\frac{\lambda '}{r}\right)+\Lambda, \nonumber\\
\kappa ^2 \left(p_r+\theta _1^{~1}\right)&=&-\frac{1}{r^2}+e^{-\lambda } \left(\frac{1}{r^2}
+\frac{\nu '}{r}\right)+\Lambda, \nonumber\\
\kappa ^2 \left(p_{\bot}+\theta _2^{~2}\right)&=&\frac{1}{4} e^{-\lambda } \left(-\lambda ' \nu '
+2 \nu ''+\left(\nu '\right)^2+\frac{2 \left(\nu '-\lambda '\right)}{r}\right)\nonumber\\
&&+\Lambda.\label{eqt}
\end{eqnarray}

We further consider the seed source for perfect fluid $T_{\mu\nu}$ (i.e. $\theta_{\mu\nu}=0$ in Eq.\eqref{eqeins}) and assume that the new metric ansatz takes the following form:
\begin{eqnarray}
\text{ds}^2=-e^{\xi\left(r\right)}\text{dt}^2+e^{\mu\left(r\right)}\text{dr}^2+r^2\text{d$\Omega $}^2,\label{seedmetric}
\end{eqnarray}
which implies the form of new Einstein field equations
\begin{eqnarray}
&&{\kappa ^2}\rho = \frac{1}{{{r^2}}} - {e^{ - \mu }}(\frac{1}{{{r^2}}} - \frac{{\mu '}}{r})-\Lambda,\nonumber\\
&&{\kappa ^2}p_r =-\frac{1}{{{r^2}}}+{e^{ - \mu }}(\frac{1}{{{r^2}}} + \frac{{\xi '}}{r})+\Lambda,\nonumber\\
&&{\kappa ^2}p_{\bot} =  - \frac{{{e^{ - \mu }}}}{4}(2\xi '' + \xi {'^2} - \mu '\xi '
+ 2\frac{{\xi ' - \mu '}}{r})+\Lambda. \label{seedeom}
\end{eqnarray}

To obtain information about the additional source $\theta_{\mu}^{~\nu}$, we must apply some distortion of this metric, namely \cite{Ovalle:2017fgl}
\begin{eqnarray}
&&\xi \to \nu =\xi+\alpha * g\left(r\right),\nonumber\\
&&e^{-\mu }\to e^{-\lambda }=e^{-\mu }+\alpha * f\left(r\right).\label{lambdamu}
\end{eqnarray}
With two Einstein field equations (Eqs.~\eqref{eqt}\eqref{seedeom}) and the metric deformation (Eq.~\eqref{lambdamu}),
the quasi-Einstein field equations for $\theta_{\mu}^{~\nu}$ can be obtained as
\begin{eqnarray}
&&\kappa ^2 \theta _0^{~0} =\alpha \frac{f}{r^2}+\alpha\frac{ f'}{r},\nonumber\\
&&\kappa ^2 \theta _1^{~1}=\alpha  f \left(\frac{1}{r^2}+\frac{\nu '}{r}\right)+\alpha  \gamma _1,\nonumber\\
&&\kappa ^2 \theta _2^{~2}=\alpha \frac{f}{4}\left(2 \nu ''+\nu '^2+\frac{2 \nu '}{r}\right)
+\alpha \frac{f'}{4} \left(\nu '+\frac{2}{r}\right)+\alpha\gamma_2,\label{thetaeq1}
\end{eqnarray}
with
\begin{eqnarray}
{\gamma _1}{\rm{ = }}\frac{{{e^{ - \mu }}g'}}{r},\quad
{\gamma _2} = \frac{{{e^{ - \mu }}}}{4}(2g'' + g{'^2} + \frac{{2g'}}{r} + 2\xi 'g' - \mu 'g').\nonumber
\end{eqnarray}

Clearly, the two sources $T_{\mu\nu}$ and $\theta_{\mu\nu}$ have been successfully decoupled  using the EGD method.

Considering the conservation equations for the total energy-momentum tensor $\tilde{T}^{\mu\nu}$,
the tensors $T^{\mu\nu}$ and $\theta^{\mu \nu }$ satisfy
\begin{eqnarray}
{\nabla _\nu }{T^{\mu \nu }} = 0,\quad
{\nabla _\nu }{{\theta}^{\mu \nu }} = 0,
\end{eqnarray}
because the tensors $T_{\mu\nu}$ and $\theta_{\mu\nu}$ are non-interacting.
Then, we can obtain the conservation equation for additional energy-momentum tensor $\theta_{\mu\nu}$
\begin{eqnarray}
(\theta_1^{~1})'-\frac{{\nu'}}{2}(\theta_0^{~0}-\theta_1^{~1})-\frac{2}{r}(\theta_2^{~2}-\theta_1^{~1})
-\frac{{\alpha g'}}{2}(T_0^{~0} - T_1^{~1})=0.
\end{eqnarray}

\section{Hairy black hole solutions}\label{bh}

Now we construct hairy black holes. For simplify, we can assume the energy-momentum $T_\mu^{~\nu}$
(seed source) to be satisfy $T_\mu^{~\nu}=0$ namely, $\rho=p_r=p_{_\bot}=0$.
From the Einstein field equations (Eq.\eqref{seedeom}),
we can easily obtain the analytical solutions for the seed metric,
\begin{eqnarray}
{e^\xi}={e^{-\mu}}=1-\frac{{2M}}{r}-\frac{{{r^2}\Lambda}}{3},\label{seedsolution}
\end{eqnarray}
which are exactly Schwarzschild AdS black hole solution.
Substituting the above metric solution into Eq.~\eqref{lambdamu},
we can obtain
\begin{eqnarray}
&&{e^{ - \xi (r)}} \to {e^{ - \nu (r)}} = (1-\frac{{2M}}{r}-\frac{{{r^2}\Lambda }}{3})*{e^{\alpha g(r)}},\nonumber\\
&&{e^{ - \mu (r)}} \to {e^{ - \lambda (r)}} = (1-\frac{{2M}}{r}-\frac{{{r^2}\Lambda }}{3}) + \alpha f(r).
\end{eqnarray}
It's worthy to point out that in order to have black hole solutions with a well defined horizon structure,
we set
\begin{eqnarray}
{e^{\nu}} = {e^{-\lambda}}.\label{nulambda}
\end{eqnarray}
Then, the killing horizon and causal horizon of the hairy black hole would be in the same position
\begin{eqnarray}
{e^{\nu (r_h)}} = {e^{ - \lambda (r_h)}} = 0.
\end{eqnarray}

From Eqs.\eqref{lambdamu} and \eqref{nulambda}, the metric deformations $f(r)$ and $g(r)$ can be rewritten as
\begin{eqnarray}
\alpha f(r)=(1-\frac{{2M}}{r}-\frac{{{r^2}\Lambda }}{3})*({e^{\alpha g(r)}} - 1).
\end{eqnarray}
The metric line element of hairy black hole becomes
\begin{eqnarray}
d{s^2}=&&-(1-\frac{{2M}}{r}-\frac{{{r^2}\Lambda}}{3})*{e^{\alpha g(r)}}\nonumber\\
&&+{(1-\frac{{2M}}{r}
-\frac{{{r^2}\Lambda}}{3})^{-1}}*{e^{-\alpha g(r)}}+{r^2}d{\Omega^2}.\label{finalmetric}
\end{eqnarray}
Therefore, the event horizon $r_h$ of this hairy black hole is given by
\begin{eqnarray}
{r_h}=&&-\frac{1}{{\left({3M{\Lambda ^2}+\sqrt{9{M^2}\Lambda^4-\Lambda ^3} } \right)}^{1/3}}\nonumber\\
&&-\frac{{\left({3M{\Lambda^2}+\sqrt{9{M^2}\Lambda^4-\Lambda ^3} } \right)}^{1/3}}{\Lambda}.\label{rh}
\end{eqnarray}
Note that the horizon radius $r_h>0$ of hairy black hole is always satisfied on account
of positive mass ($M>0$) of black hole and negative cosmological constant ($\Lambda<0$).

According to the new metric ansatz of hairy black hole \eqref{finalmetric} and the seed solutions \eqref{seedsolution},
the quasi-Einstein field equations \eqref{thetaeq1} for $\theta_{\mu}^{~\nu}$ become
\begin{eqnarray}
\kappa ^2 \theta _0^{~0} =&&\Lambda-\frac{1}{r^2}+e^{\alpha g(r)}\left(\Lambda-\frac{1}{r^2}
+\alpha \left(\frac{1}{r}-\frac{2M}{r^2}-\frac{\Lambda r}{3}\right)\right)g(r),\nonumber\\
\theta _1^{~1}=&&\theta _0^{~0},\nonumber\\
\kappa ^2 \theta _2^{~2}=&&\Lambda+e^{\alpha g(r)}\Big[\alpha\left(\frac{1}{2}-\frac{M}{r}
-\frac{\Lambda r^2}{6}\right)\left(\alpha g'(r)^2+g''(r)\right)\nonumber\\
&&+\alpha\left(\Lambda r-\frac{1}{r}\right)g'(r)-\Lambda\Big].\label{thetaeq2}
\end{eqnarray}
It is clear that the deformation metric $g(r)$ is only determined by the additional source $\theta _\mu^{~\nu}$.
We can further derive the quasi-Einstein system to obtain the solutions of hairy black holes
by assuming that the source $\theta _\mu^{~\nu}$ satisfies
some physically motivated equations of state.
We first consider the simple case of isotropic pressure
\begin{eqnarray}
\theta _0^{~0}=\theta _1^{~1}=\theta _2^{~2}.
\end{eqnarray}
Eq.~\eqref{thetaeq2} then yields a differential equation
\begin{eqnarray}
6(h(r)-1)+4(r^3\Lambda-3M)h'(r)+(6 M r+r^4\Lambda-3 r^2)h''(r)=0,
\end{eqnarray}
where the function ${\rm{h}}(r)$ denotes ${e^{\alpha g(r)}}$.
Then, we have
\begin{eqnarray}
{\rm{h}}(r)={e^{\alpha g(r)}}=1+\frac{C_1+ r^3 C_2}{{6M-3r + {r^3}\Lambda }}, \label{isosoluh}
\end{eqnarray}
where $C_1$ and $C_2$ are integration constants.
Finally, we obtain the solution of hairy black hole
\begin{eqnarray}
{e^\nu}={e^{-\lambda }}=1-\frac{2M}{r}-\frac{{{r^2}\Lambda}}{3}-\frac{{{C_1}}}{3r}
-\frac{C_2 r^2}{9}.
\end{eqnarray}

Note that it takes the same form as the solution of Schwarzchild AdS black hole by
adopting the redefinition of the asymptotic mass $M\rightarrow M=M+C_1/6$
and cosmological constant $\Lambda\rightarrow \Lambda=\Lambda+C_2/3$.

Moreover, considering the seed source $T_{\mu\nu}=0$ previously,
the effective density, effective radial pressure and effective tangential pressure
are obtained as
\begin{eqnarray}
\tilde{\rho}=-\theta_0^{~0}=\frac{C_2}{3},\quad \tilde{p}_r=\tilde{p}_{_\bot}=-\frac{C_2}{3}.
\end{eqnarray}
which implies it is exactly isotropic fluid.

Now, let us consider the linear equation of state and assume that the components of source $\theta _\mu^{~\nu}$
satisfy
\begin{eqnarray}
\theta _0^{~0}= a\theta _1^{~1} + b\theta _2^{~2},
\end{eqnarray}
where $a$ and $b$ are arbitrary constants.
From the quasi-Einstein field equations (Eq.\eqref{thetaeq2}),
we can easily obtain
\begin{eqnarray}
&&6[a-1-(a+b-1){r^2}\Lambda]+6[1-a+(a+b-1){r^2}\Lambda ]h(r)\nonumber\\
&&+2[6(a-1)M-3(a+b-1)r+(a+3b-1){r^3}\Lambda]h'(r) \nonumber\\
&&+br(6M-3r+{r^3}\Lambda )h''(r)=0.
\end{eqnarray}
Then, the solution can be written as
\begin{eqnarray}
{e^{\alpha g(r)}}=h(r) = \frac{{-3r + {r^3}\Lambda  + {C_1} + \frac{{b{r^{\frac{{2 - 2a
+ b}}{b}}}{C_2}}}{{2 - 2a + b}}}}{{6M-3r + {r^3}\Lambda }}.\label{linearsolu}
\end{eqnarray}
Here, $C_1$ and $C_2$ are constants with dimensions
of length.

Considering the line element of the hairy black hole \eqref{finalmetric}, we obtain the solution of the hairy black hole
\begin{eqnarray}
{e^\nu}={e^{-\lambda }}=1-\frac{{{r^2}\Lambda}}{3}-\frac{{{C_1}}}{3r}
-\frac{{b{r^{\frac{{2-2a}}{b}}}{C_2}}}{6-6a+3b}.
\end{eqnarray}
Because $T_{\mu\nu}=0$ was defined previously, the corresponding components of $\theta_{\mu\nu}$ are obtained as the effective density
\begin{eqnarray}
\tilde{\rho}=-\theta_0^{~0}=-\frac{1}{3}{C_2}{r^{-\frac{{2\left({-1+a+b}\right)}}{b}}},
\end{eqnarray}
the effective radial pressure
\begin{eqnarray}
\tilde{p}_r=\theta_1^{~1}=\theta _0^{~0}=-\rho,
\end{eqnarray}
and the effective tangential pressure
\begin{eqnarray}
\tilde{p}_{_\bot}=\theta _2^{~2}=-\frac{{\left({-1 + a}\right)C_2{r^{-\frac{{2\left({-1+a+b}\right)}}{b}}}}}{3b}.
\end{eqnarray}
The anisotropy is thus given by
\begin{eqnarray}
\Pi=\tilde{p}_r-\tilde{p}_{_\bot}=-\frac{{\left({-1+a+b}\right)C_2{r^{-\frac{{2\left({-1+a+b}\right)}}{b}}}}}{3b} \ne 0.\label{linearsol}
\end{eqnarray}

If setting $n=\frac{{2-2a}}{b}$ and $C_1=6M$, this solution \eqref{linearsol} reduces to
the Kiselev-AdS black hole with quintessence. In addition,
we have checked some other scenarios when the additional source $\theta_{\mu}^{~\mu}$ satisfies the conformal equations
of state $\theta_{\mu}^{~\mu}=0$ or the polytropic $\alpha\theta_{1}^{~1}=K\left(\alpha\theta_{0}^{~0}\right)^\Gamma$.
However, these hairy black hole solutions have already been
studied in Refs.\cite{Kiselev:2002dx}-\cite{Rodrigues:2022qdp}.

It is worth pointing out that the energy conditions are generally considered as sensible guidelines to avoid classically
unphysical configurations. In Ref.~\cite{Curiel:2014zba}, energy conditions are usually imposed on the energy-momentum
tensor to avoid exotic matter sources in the context of a wide class of spacetime theories.
In the next sections, we discuss black holes in asymptotical AdS spacetime
when energy conditions are imposed on the source $\theta_{\mu\nu}$.

\subsection{Strong energy condition}

In energy condition theory, there is an important condition known as the SEC,
which arises from proving the singularity theorem. It can be written as follows\cite{liang}:
\begin{eqnarray}
\tilde{\rho} + \tilde{p}_r + 2\tilde{p}_{_\bot } \ge 0,\label{sec1} \\
\tilde{\rho} + \tilde{p}_r \ge 0, \label{sec2}\\
\tilde{\rho} + \tilde{p}_{_\bot } \ge 0.\label{sec3}
\end{eqnarray}
Considering the choice for metric ansatz ${e^{\nu}} = {e^{-\lambda}}$ \eqref{nulambda},
Eqs.~\eqref{sec1}-\eqref{sec3} become
\begin{eqnarray}
&&\theta _2^{~2} \ge 0, \label{th1}\\
&&\theta _0^{~0} \le \theta _2^{~2}.\label{th2}
\end{eqnarray}

According to the quasi-Einstein field equations for additional source $\theta_{\mu}^{~\nu}$ (Eq.\eqref{thetaeq2}),
conditions \eqref{th1} and \eqref{th2} lead to the inequalities
\begin{eqnarray}
{G_1}(r)&\equiv&6h'(r)-6Mh''(r)+3rh''(r)-\Lambda\Big(-6r+6rh(r)\nonumber\\
&&+6{r^2}h'(r)+{r^3}h''(r)\Big)\ge 0 \label{g1r},\\
{G_2}(r)&\equiv&6-6h(r)+12Mh'(r)-6Mrh''(r)\nonumber\\
&&+3{r^2}h''(r)-\Lambda(4{r^3}h'(r)+{r^4}h''(r))\ge 0\label{g2r},
\end{eqnarray}
where the function ${\rm{h}}(r)$ denotes ${e^{\alpha g(r)}}$.

Let us first compute the corresponding boundary solution of the above two expressions.
From ${G_1(r)=0}$, one can easily obtain
\begin{eqnarray}
h(r) = \frac{{{r^3}\Lambda + C_{11}+ r C_{12} }}{{6M-3r+{r^3}\Lambda}}.
\end{eqnarray}
Then, we can obtain the black hole solution from the metric \eqref{finalmetric}
\begin{eqnarray}
{{\rm{e}}^\nu } = {e^{ - \lambda }} =-\frac{{{r^2}\Lambda }}{3}-\frac{1}{3}( \frac{{{C_{11}}}}{r} + {C_{12}}).\label{h11r}
\end{eqnarray}
For the inequality of ${G_2(r)}$ \eqref{g2r}, the corresponding boundary solution can be also obtained from ${G_2(r)=0}$
\begin{eqnarray}
h(r) = 1 + \frac{{{C_{21}} + \frac{1}{3}{r^3}{C_{22}}}}{{6M-3r+{r^3}\Lambda }},
\end{eqnarray}
and
\begin{eqnarray}
{{\rm{e}}^\nu } = {e^{ - \lambda }} = 1 - \frac{{2M}}{r}-\frac{{{r^2}\Lambda }}{3}
-\frac{{{C_{21}}}}{3r}-\frac{1}{9} {r^2}{C_{22}}.\label{h22r}
\end{eqnarray}
To agree on the Eq.\eqref{h11r} and Eq.\eqref{h22r}, we can set $C_{11}=C_{21}+6M$,
$C_{12}=-3$ and $C_{22}=0$. Then the metric becomes
\begin{eqnarray}
{{\rm{e}}^\nu}={e^{-\lambda }}=1 - \frac{{2M}}{r}-\frac{{{r^2}\Lambda }}{3}-\frac{{{C_{21}}}}{3r}.
\end{eqnarray}
This is actually the Schwarzschild AdS black hole. In other words,
the black hole solution derived from the boundary conditions is mediocre.

Now let us look for a nontrivial solution of function $h(r)$.
From Eq.~\eqref{g1r}, we could choose a simple model, i.e
\begin{eqnarray}
G_1(r)\equiv{G_\alpha (r)}=\frac{\alpha }{A_1}(r-2M-\frac{{{r^3}\Lambda}}{3}){e^{\frac{{-r}}{B_1}}},\label{ga}
\end{eqnarray}
where $A_1$ and $B_1$ are pending parameters.
The function $G_{\alpha} (r)$ satisfies clearly
${G_1}(r) = 0$ near the horizon $r = r_h$ and at infinity $r \to \infty$ and also meets
\begin{eqnarray}
{G_1}(r)>0, \quad  r_h<r<\infty
\end{eqnarray}
when $A_1>0$.
After substituting Eq.~\eqref{ga} into Eq.~\eqref{g1r}, we get
\begin{eqnarray}
h(r)&=&\frac{1}{3A_1(6M-3r+{r^3}\Lambda )}\Big({B_1^2\alpha e^{-\frac{r}{B_1}}}\left(24{B_1^3}\Lambda+6M-3r+{r^3}\Lambda\right.\nonumber\\
&&\left. +18{B_1^2}r+6B_1({r^2}\Lambda-1)\right)+3{A_1}({r^3}\Lambda+{C_1}+r{C_2})\Big).\label{hrsec}
\end{eqnarray}
Then, the solution of hairy black hole is obtained as
\begin{eqnarray}
e^\nu={e^{-\lambda}}&=&-\frac{{{r^2}\Lambda}}{3}-\frac{{{C_1}+r{C_2}}}{{3r}}
+{e^{-\frac{r}{B_1}}}\alpha\left(\frac{{B_1^2}(-2M+2B_1+r)}{3{A_1}r}\right.\nonumber\\
&&\left.- \frac{{(24{B_1^5} + 18{B_1^4}r + 6{B_1^3}{r^2} + {B_1^2}{r^3})\Lambda }}{{9{A_1}r}}\right)\label{eqe1}.
\end{eqnarray}
If setting $A_1=B_1^2$, Eq.(\ref{eqe1}) reduces to
\begin{eqnarray}
{e^\nu }={e^{-\lambda}}&=&{e^{-\frac{r}{B_1}}}\alpha\Big(\frac{-2M+2B_1+r}{3r}-\frac{{(24{B_1^3}+18{B_1^2}r+6B_1{r^2}+{r^3})\Lambda }}{{9r}}\Big)\nonumber\\
&&-\frac{{{r^2}\Lambda}}{3}-\frac{1}{3}(\frac{{{C_1}}}{r}+{C_2})\label{eqe2}.
\end{eqnarray}

Furthermore, if we continue to set $ B_1=M$, $C_1=6M$, $C_2=-3$, Eq.(\ref{eqe2}) reduces to
\begin{eqnarray}
{e^\nu}={e^{-\lambda }}&=&1-\frac{2M}{r}-\frac{1}{3}\alpha\Big(\frac{(24M^3+18M^2r+6Mr^2+r^3)\Lambda}{3r} \nonumber\\
&&+1\Big)e^{-\frac{r}{M}}-\frac{r^2\Lambda }{3}.\label{linesec}
\end{eqnarray}

\begin{figure}
\begin{minipage}[t]{0.5\linewidth}
\centering
\includegraphics[width=5.6cm,height=4.4cm]{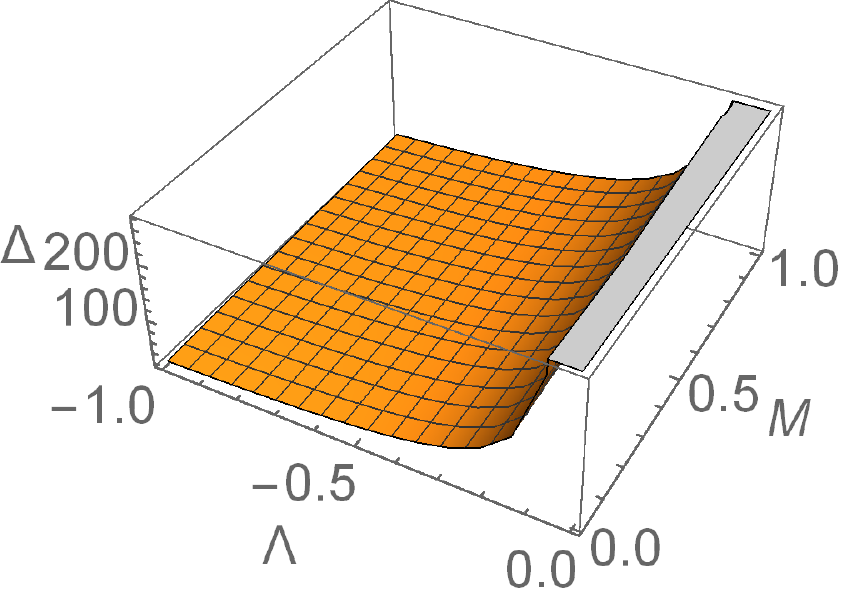}
\hfill%
\end{minipage}%
\begin{minipage}[t]{0.5\linewidth}
\centering
\includegraphics[width=4.6cm,height=4.4cm]{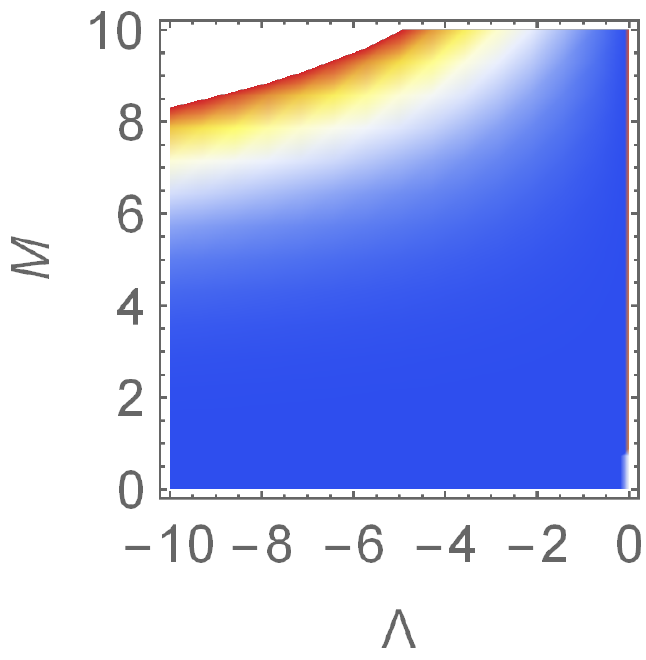}
\includegraphics[width=1.7cm,height=4.4cm]{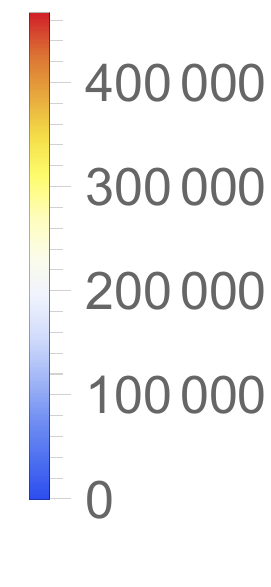}
\hfill%
\end{minipage}
\caption{The value of $\Delta$ at the horizon. The left graph is the surface diagram of $\Delta$ with respect
to $\Lambda$ and $M$, the corresponding values of $\Lambda$ and $M$ are respectively $-1$ to $0$ and $0$ to $1$.
The right graph is the density map of $\Delta$ and the value range of $\Lambda$ and $M$ are respectively $-10$ to $0$ and $0$ to $10$.}\label{fig1}
\end{figure}

In addition, we must check that the function $h(r)$ satisfies the inequality Eq.(\ref{g2r}). Substitute Eq.\eqref{hrsec} to Eq.(\ref{g2r}), we can get
\begin{eqnarray}
 \Delta=&&-6{M^2}+3{r^2}-(12{M^4} + 12{M^3}r + 6{M^2}{r^2}\nonumber\\
 &&+ 2M{r^3} + {r^4})\Lambda\ge 0.
\end{eqnarray}
In AdS space-time ($\Lambda<0$), $\Delta$ is a monotonically increasing function of $r$.
Therefore, as long as the value of the function is positive at the horizon, it must also be positive
outside the horizon. Because the analytic solution of the event horizon \eqref{rh} of the black hole
is rather complicated, we use the numerical simulation method to analyze the value of $\Delta$.
The behavior of this function at the event horizon of the black hole is plotted in Fig.\ref{fig1}.
The figures show that no matter how the values of $\Lambda$ and $M$ change, $\Delta$ is always positive.
Therefore, the inequality Eq.(\ref{g2r}) is always true at $r \ge r_h$.

For the line element in Eq.(\ref{linesec}), we can also find its effective density
\begin{eqnarray}
-\tilde{\rho} = \theta _0^{~0}=\tilde{p}_r=&&\frac{{{{\rm{e}}^{ - \frac{r}{M}}}\alpha }}{{9M{r^2}}}(3M-3r+6{M^3}\Lambda+6{M^2}r\Lambda \nonumber\\
&&+3Mr^2\Lambda  + r^3\Lambda ),\label{rhoL}
\end{eqnarray}
and the effective tangential pressure
\begin{eqnarray}
\tilde{p}_{_\bot }=\theta_2^{~2}=\frac{{{{\rm{e}}^{-\frac{r}{M}}}\alpha\left({-6M+3r -{r^3}\Lambda}\right)}}{{18{M^2}r}}.\label{pL}
\end{eqnarray}
Note that when $\Lambda\to 0$, Eqs.\eqref{linesec}, \eqref{rhoL} and \eqref{pL}
degenerate to the results in Ref. \cite{Ovalle:2018umz}.

\subsection{Dominant energy condition}

Now let us consider another condition. The DEC requires that the four-momentum density
measured by any instantaneous viewer is a time-like or light-like vector, whose physical interpretation
is that the energy flow rate of the material field is less than or equal to the speed of light, which is equivalent to \cite{liang}
\begin{eqnarray}
&&\tilde \rho  \ge \left| {{{\tilde p}_r}} \right|,\\
&&\tilde \rho  \ge \left| {{{\tilde p}_t}} \right|,
\end{eqnarray}
in this system of Eq.~\eqref{finalmetric}. These two inequalities reduces to
\begin{eqnarray}
\begin{array}{l}
-\theta _0^{~0} - \theta _2^{~2} \ge 0,\\
-\theta _0^{~0} + \theta _2^{~2} \ge 0,
\end{array}
\end{eqnarray}
which yield respectively the differential inequalities
\begin{eqnarray}
{H_{\rm{1}}}(r) & = &2 - 2h(r) + \left( {4M - 4r} \right)h'(r) + \left( {2Mr - {r^2}} \right){h^{\prime \prime }}(r)\nonumber\\
 &&+ \Lambda \left( {4{r^2} - 4{r^2}h(r) + \frac{8}{3}{r^3}h'(r) + \frac{1}{3}{r^4}{h^{\prime \prime }}(r)} \right)\ge 0,\label{H1r}\\
{H_{\rm{2}}}(r) & = & 2 - 2h(r) + 4Mh'(r) + \left( { - 2Mr + {r^2}} \right){h^{\prime \prime }}(r)\nonumber\\
&& - \Lambda \left( {\frac{4}{3}{r^3}h'(r) + \frac{1}{3}{r^4}{h^{\prime \prime }}(r)} \right)\ge 0,\label{H2r}
\end{eqnarray}
respectively, for which we used Eq.\eqref{thetaeq2} again. Now, choosing the same model as for the SEC (Eq.(\ref{ga})),
\begin{eqnarray}
H_1(r)\equiv{H_\alpha(r)}=\frac{\alpha }{A_1}(r-2M-\frac{{{r^3}\Lambda }}{3}){e^{\frac{{-r}}{B_1}}},\label{ha}
\end{eqnarray}
we can solve the Eq.\eqref{H1r} and obtain
\begin{eqnarray}
h(r)&=&\frac{\rm{e}^{-\frac{r}{{{B_1}}}}}{{r({6M-3r+{r^3}\Lambda }){A_1}}}\Big((-\alpha B_1^2(6M-3r+{r^3}\Lambda+6(-1+{r^2}\Lambda){B_1}\nonumber\\
&&+18r\Lambda B_1^2+24\Lambda B_1^3)){\rm{+}}(-3{r^2}+{r^4}\Lambda+{C_1}+r{C_2}){A_1}\Big).
\end{eqnarray}
Similarly, let us take the same approach as Eq.\eqref{eqe2}, that is $A = B^2$,
\begin{eqnarray}
h(r)&=&\frac{\rm{e}^{-\frac{r}{B_1}}}{r({6M-3r+{r^3}\Lambda })}\Big(-\alpha(6M-3r+{r^3}\Lambda+6({-1+{r^2}\Lambda}){B_1}+18r\Lambda B_1^2\nonumber\\
&& + 24\Lambda B_1^3) + (-3{r^2} + {r^4}\Lambda  + C_1 + r C_2 )\Big).\label{hrdec}
\end{eqnarray}
Furthermore, setting $B = M$, $C_1 = - 3 Q^2$, $C_2 =6M$, the metric finally can be written as
\begin{eqnarray}
e^\nu = e^{ - \lambda } & = & \alpha {{\rm{e}}^{ - \frac{r}{M}}}\left( { - \frac{1}{r}
- \frac{{\left( { - 24{M^3} - 18{M^2}r - 6M{r^2} - {r^3}} \right)\Lambda }}{{3{r^2}}}} \right)\nonumber\\&&
+ 1 - \frac{{2M}}{r} + \frac{{{Q^2}}}{{{r^2}}} - \frac{{{r^2}\Lambda }}{3}.
\end{eqnarray}
As with the SEC, we must check whether $h(r)$ \eqref{hrdec} satisfies the other inequality for the DEC \eqref{H2r}, and we get the inequality
\begin{eqnarray}
&&\frac{{4{Q^2}}}{{{r^2}}} - {{\rm{e}}^{ - \frac{r}{M}}}\alpha \frac{1}{{3{r^2}}}\Big( \frac{{6{r^2}}}{M}
+ \frac{{3{r^3}}}{{{M^2}}} - ( 96{M^3} + 96{M^2}r + 48M{r^2} \nonumber\\
&&+ 16{r^3} + \frac{{4{r^4}}}{M} + \frac{{r^5}}{{M^2}} )\Lambda  \Big) \ge 0.\label{checkdec}
\end{eqnarray}
The inequality Eq.\eqref{checkdec} is not guaranteed to be satisfied at any point,
the necessary condition is that
\begin{eqnarray}
\frac{{12{Q^2}}}{\alpha }\ge &&\frac{{6{r^2}}}{M}+\frac{{3{r^3}}}{{{M^2}}}-\left( 96{M^3}
+ 96{M^2}r + 48M{r^2}\right.\nonumber\\
 &&\left.+ 16{r^3} + \frac{{4{r^4}}}{M} + \frac{{{r^5}}}{{{M^2}}} \right)\Lambda
\end{eqnarray}

On the premise that the above condition is satisfied, if we treat $Q$ as the electric charge, it is clear that
this black hole is a deformation of the charged black hole in asymptotically AdS spacetime.
Moreover, the effective density can be written as
\begin{eqnarray}
\tilde{p}_r =   \theta _1^{~1} & = & \frac{1}{{{r^4}}}\Big( - (24{{\rm{e}}^{ - \frac{r}{M}}}{M^3}\alpha
+ 24{{\rm{e}}^{ - \frac{r}{M}}}{M^2}r\alpha  + 12{{\rm{e}}^{ - \frac{r}{M}}}M{r^2}\alpha \nonumber\\&&
+ 4{{\rm{e}}^{ - \frac{r}{M}}}{r^3}\alpha  + \frac{{{{\rm{e}}^{ - \frac{r}{M}}}{r^4}\alpha }}{M})\Lambda
+ \frac{{{{\rm{e}}^{ - \frac{r}{M}}}{r^2}\alpha }}{M} - {Q^2}\Big),
\end{eqnarray}
the effective tangential pressure is
\begin{eqnarray}
\tilde{p}_{_\bot} = \theta_2^{~2}&=&\frac{1}{{{r^4}}}\Big(\frac{1}{6}(48{{\rm{e}}^{ - \frac{r}{M}}}{M^3}\alpha
+ 48{{\rm{e}}^{ - \frac{r}{M}}}{M^2}r\alpha  + 24{{\rm{e}}^{ - \frac{r}{M}}}M{r^2}\alpha + 8{{\rm{e}}^{ - \frac{r}{M}}}{r^3}\alpha  \nonumber\\&&
+ \frac{{2{{\rm{e}}^{ - \frac{r}{M}}}{r^4}\alpha }}{M} + \frac{{{{\rm{e}}^{ - \frac{r}{M}}}{r^5}\alpha }}{{{M^2}}}\Lambda )
+ {Q^2} - \frac{{{{\rm{e}}^{ - \frac{r}{M}}}{r^3}\alpha }}{{2{M^2}}}\Big).
\end{eqnarray}
Then, we can obtain
\begin{eqnarray}
\tilde{p}_r - \tilde{p}_{_ \bot }=&&\frac{1}{{r^4}}\Bigg(\frac{{{{\rm{e}}^{ - \frac{r}{M}}}{r^2}\left( {2M + r} \right)\alpha }}{{2{M^2}}}
- 2{Q^2} - \frac{1}{6}\Big(48{{\rm{e}}^{ - \frac{r}{M}}}{M^2}\left( {2M + r} \right)\alpha \nonumber\\&&
+ 24{{\rm{e}}^{ - \frac{r}{M}}}Mr\left( {2M + r} \right)\alpha  + 12{{\rm{e}}^{ - \frac{r}{M}}}{r^2}\left( {2M + r} \right)\alpha \nonumber\\&&
+ \frac{{2{{\rm{e}}^{ - \frac{r}{M}}}{r^3}\left( {2M + r} \right)\alpha }}{M} + \frac{{{{\rm{e}}^{ - \frac{r}{M}}}{r^4}\left( {2M
+ r} \right)\alpha }}{{{M^2}}}\Big)\Lambda \Bigg).
\end{eqnarray}
Therefore, the source $\theta_{\mu\nu}$ under dominant energy condition is also anisotropic.

\section{Conclusion and discussion}\label{con}

For two non-interacting gravitational sources in the gravitational system,
we adopt gravitational decoupling approach via EGD to present how the Schwarzschild AdS black hole
is modified when the vacuum is filled by new fields $\theta_{\mu\nu}$.
The EGD technique is devised for describing deformations of known solutions
of GR induced by adding extra sources.
We first consider several scenarios, where the source term $\theta_{\mu\nu}$ satisfies the case
of isotropic pressure or linear equation of state.
Then, we obtain the corresponding solutions of hairy black holes by derived the quasi-Einstein
field equations for additional source $\theta _\mu^{~\nu}$. However, these black hole solutions
are mediocre and have already been extensively studied in many papers.

The SEC and DEC have been adopted
as constraint conditions for the construction for hairy black hole because
the energy conditions are generally considered
as sensible guidelines to avoid the exotic matter sources
in the context of a wide class of spacetime theories.
We find two new families of hairy black holes in the asymptotical AdS spacetime.
In addition, if we regard the black hole spacetime as a fluid system,
the fluid under each conditions is anisotropic.

Finally, let us comment on some concrete prospects of this study.
Currently, black hole thermodynamics in the presence of a negative cosmological constant is becoming
even more appealing because they allow a gauge duality description through a thermal field theory via AdS/CFT correspondence~\cite{Witten:1998qj}.
The gravitational decoupling method has been applied to discuss the thermodynamics of spherically symmetric black hole in asymptotically
flat spacetime~\cite{Estrada:2020ptc}, where the first law of thermodynamics was decoupled into two sectors:
the standard first law of thermodynamics and quasi first law of thermodynamics. Therefore, it's interesting to investigate
the thermodynamics and phase transitions of hairy black holes in asymptotically AdS spacetime.

On the other hand, the entanglement entropy is a versatile tool and may provide us new insights into a
rich variety of physical phenomena, ranging from the confining phase of large-$N$ gauge theories~\cite{Klebanov:2007ws}
to topological phases in condensed matter systems~\cite{Kitaev:2005dm}, and tachyon condensation~\cite{Nishioka:2006gr}
in the light of the AdS/CFT correspondence. S.~Ryu and T.~Takayanagi~\cite{Ryu:2006bv,Nishioka:2009un} proposed
a holographic description of the entanglement entropy in quantum (conformal) field theories
via AdS$_{d+2}$/CFT$_{d+1}$ correspondence, and computed the entanglement entropy at finite
temperature employing AdS black hole geometry. Then, it's worth discussing the holographic entanglement entropy (HEE)
for the hairy AdS black holes obtained by using EGD approach.
In addition, the stability and quasinormal modes (QNMs) of hairy black holes caused by gravitational decoupling
under various field perturbations have been investigated in Refs.~\cite{Cavalcanti:2022cga,Yang:2022ifo}.
The QNMs play a fundamental role in characterizing gravitational
wave signals detected by LIGO and VIRGO~\cite{LIGOScientific:2018dkp,LIGOScientific:2019fpa}.
We will investigate the stability and QNMs of hairy AdS black holes in future work.
The final results could be expected to provide some direction for observing hairy
black holes caused by gravitational decoupling in future experiments.

 \vspace{1cm}

{\bf Acknowledgments}

We appreciate Xiao-Mei Kuang and Sheng-Yuan Li for helpful discussions. D. C. Zou acknowledges financial support from
Outstanding young teacher programme from Yangzhou University, No. 137050368.

\end{document}